\documentclass[12pt,english]{article}
\usepackage[latin9]{inputenc}
\usepackage{geometry}
\usepackage{amssymb}
\usepackage{graphicx}
\usepackage{setspace}
\makeatletter
\@ifundefined{date}{}{\date{}}
\makeatother

\usepackage{babel}
\begin{document}

\title{Quantum Analog of Carnot Engine}

\author{{\normalsize{}Chung-Yang Wang}\\
\emph{\normalsize{}Department of Physics, National Taiwan University,
Taipei 106, Taiwan, ROC}\\
{\normalsize{}Email: r03222014@ntu.edu.tw}}
\maketitle
\begin{abstract}
In order to build a quantum analog of traditional Carnot engine, a
common choice is replacing the two thermodynamic adiabatic processes
with two quantum mechanical adiabatic processes. In general, such
quantum Carnot engine has six strokes. We analyze the efficiency of
such six-stroke quantum Carnot engine in a perturbative way. The analytic
analysis matches the numerical result in \cite{xiao2015construction}.
\end{abstract}
\medskip{}

\medskip{}
\medskip{}

\medskip{}
\medskip{}

\medskip{}

\medskip{}
\medskip{}

\medskip{}

\medskip{}

\medskip{}
\medskip{}

\medskip{}

\medskip{}

\medskip{}

\section{Introduction}

In classical thermodynamics, a Carnot engine consists of two isothermal
processes and two thermodynamic adiabatic processes. When trying to
establish a quantum mechanical analog of traditional Carnot engine,
we identify the counterparts of these thermodynamic processes, and
then patch them to build a quantum Carnot engine (abbreviated as QCE)
\cite{bender2000quantum,quan2007quantum}.

So how to establish these counterparts? For isothermal process, we
identify quantum mechanical isothermal process in a statistical mechanical
way by requiring the populations of the substance to satisfy Boltzmann
distribution with a fixed temperature \cite{quan2007quantum}. As
for adiabatic process, a common choice is replacing thermodynamic
adiabatic process ($dQ=\sum_{n}E_{n}dP_{n}=0$) with quantum mechanical
adiabatic process ($dP_{n}=0$) \cite{abah2014efficiency,quan2007quantum}.
We want to emphasize that a quantum mechanical adiabatic process is
conceptually different from a thermodynamic adiabatic process. This
fundamental difference is the origin of the differences between traditional
Carnot engine and QCE, as we will see in the following.

In classical thermodynamics, if a system is initially at thermal equilibrium,
it will still be at thermal equilibrium at the end of a thermodynamic
adiabatic process. On the other hand, quantum mechanical adiabatic
process doesn't enjoy such property. Consider a system undergoes a
quantum mechanical adiabatic process from state $G$ (at thermal equilibrium
with temperature $T_{G}$) to state $H$. For populations, we have
$P_{n}(G)=P_{n}(H)$ for all energy levels. Condition for state $H$
being at thermal equilibrium with temperature $T_{H}$ is

\begin{equation}
E_{n}(H)-E_{m}(H)=\frac{T_{H}}{T_{G}}\left(E_{n}(G)-E_{m}(G)\right),\label{eq:B11}
\end{equation}
for all $n$ and $m$. Since Eq.(\ref{eq:B11}) is not always satisfied,
quantum mechanical adiabatic process doesn't necessarily bring a thermal
equilibrium state to another thermal equilibrium state.

Thermal equilibrium is preserved if Eq.(\ref{eq:B11}) is respected,
that is, scale invariance is satisfied. Examples of this kind include
a harmonic oscillator with the varied parameter being the harmonic
frequency \cite{abah2014efficiency,rossnagel2014nanoscale}, a particle
in an infinite square well potential with the varied parameter being
the width of the potential well \cite{bender2000quantum}.

If Eq.(\ref{eq:B11}) is not satisfied, state $H$ is not at thermal
equilibrium. In this case, when the working medium (of state $H$)
contacts the thermal reservoir, we get an additional relaxation process
(working medium to be thermalized by the reservoir). The cycle thus
has six strokes, as shown in Fig. 1. Apparently, traditional Carnot
engine and such six-stroke QCE have many differences, including their
efficiencies. What we are going to do is studying the efficiency and
its optimization of the six-stroke QCE.

\begin{figure}[h]

\begin{centering}
\includegraphics[scale=0.65]{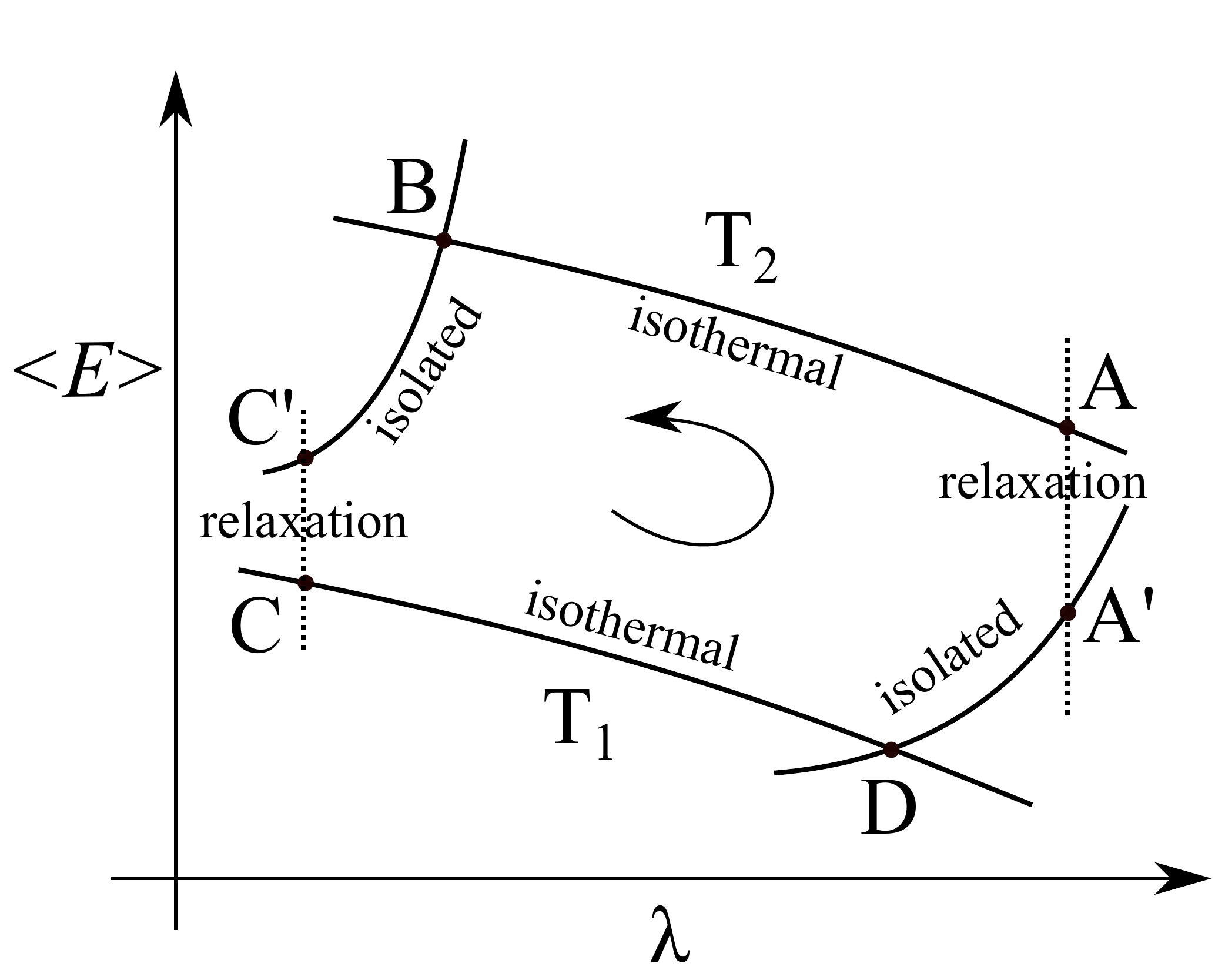}\caption{Six-stroke quantum Carnot engine. $A\rightarrow B$ and $C\rightarrow D$
are isothermal processes; $B\rightarrow C'$ and $D\rightarrow A'$
are adiabatic processes; $C'\rightarrow C$ and $A'\rightarrow A$
are relaxation processes; $T_{1}$ and $T_{2}$ are the temperatures
of the cold reservoir and the hot reservoir, respectively; $\left\langle E\right\rangle $
is the mean energy of the system. For simplicity, we assume there
is only one system parameter tunable in the cycle operation, called
$\lambda$.}
\par\end{centering}
\end{figure}

\section{Efficiency of the quantum Carnot engine and the working medium being
considered}

Following the discussion of such six-stroke QCE in \cite{xiao2015construction},
the efficiency is
\begin{equation}
\eta=1-\frac{Q_{1}}{Q_{2}}=1-\frac{T_{1}(S_{B}-S_{D})+T_{1}\Delta S_{C'\rightarrow C}^{total}}{T_{2}(S_{B}-S_{D})-T_{2}\Delta S_{A'\rightarrow A}^{total}},\label{eq:C8}
\end{equation}
where $\Delta S^{total}$ is the total entropy increase of the universe.

We can see that $\eta$ differs from Carnot efficiency, due to the
existence of the relaxation processes. By the second law of thermodynamics,
both $\Delta S_{C'\rightarrow C}^{total}$ and $\Delta S_{A'\rightarrow A}^{total}$
cannot be negative, and hence $\eta$ is lower than Carnot efficiency.
Standard Carnot engine has an universal efficiency, the Carnot efficiency,
no matter what are its working substance and the thermal reservoirs.
On the contrary, efficiency of six-stroke QCE depends on the detail
of the working substance.

In this paper, we consider six-stroke QCE whose degree of nonequilibrium
is small. That is, the scenario is the six-stroke QCE perturbed around
ordinary four-stroke Carnot engine. The spectrum of the working medium
being considered is

\begin{equation}
E_{n}=\lambda f(n)+\alpha g(n),\label{eq:D1}
\end{equation}
where $f(n)$ has no degeneracy. Here $\lambda$ is the system parameter
tuned in the cycle operations; $\alpha$ is a small constant, and
the term $\alpha g(n)$ aims to break the scale invariance and thus
lead to the two relaxation processes. The six-stroke QCE reduces to
ordinary four-stroke Carnot engine when $\alpha=0$, and $\lambda$
is denoted as $\lambda^{0}$ in this case. Our goal is maximizing
the efficiency (in a perturbative way) with $\lambda_{C}$ and $\lambda_{A}$,
while $T_{1}$, $T_{2}$, $\lambda_{B}$ and $\lambda_{D}$ are fixed.
We consider $\delta\lambda_{C}$ and $\delta\lambda_{A}$ to be of
order $\alpha$. That is,

\begin{equation}
\lambda=\lambda^{0}+\alpha\lambda^{1}.\label{eq:D2}
\end{equation}
By construction, $\lambda_{B}^{1}=\lambda_{D}^{1}=0$, $\lambda_{C}=\lambda_{C}^{0}+\alpha\lambda_{C}^{1}$,
$\lambda_{A}=\lambda_{A}^{0}+\alpha\lambda_{A}^{1}$.

Our work is inspired by \cite{xiao2015construction}. The paper considers
some general discussion, and then the numerical result of a special
case. Different from \cite{xiao2015construction}, here we analyze
systems whose spectrums are described by Eq.(\ref{eq:D1}). Since
we take an analytic approach, $\alpha$ needs to be small enough such
that perturbation can work appropriately.

\section{Calculation of heat exchange}

Let's calculate $Q_{1}$ and $Q_{2}$, with energy levels specified
by Eq.(\ref{eq:D1}). In this work, perturbation is kept to $\mathcal{O}(\alpha^{2})$.
Symbolically, the zeroth order of population and the zeroth order
of partition function are denoted as $P_{n}^{0}$ and $Z^{0}$, respectively.

For $\alpha=0$, there is no relaxation process, and hence $P_{n}^{0}(B)=P_{n}^{0}(C)$
and $P_{n}^{0}(D)=P_{n}^{0}(A)$. Furthermore, by Eq.(\ref{eq:B11}),
we obtain

\begin{equation}
\beta_{2}\lambda_{B}=\beta_{1}\lambda_{C}^{0},~\beta_{1}\lambda_{D}=\beta_{2}\lambda_{A}^{0},\label{eq:D15}
\end{equation}
where $\beta\equiv\frac{1}{k_{B}T}$. Define $\left\langle O\right\rangle $
as the average of $O_{n}$ weighted by $P_{n}^{0}$:

\begin{equation}
\left\langle O\right\rangle \equiv\sum_{n}P_{n}^{0}O_{n}.\label{eq:D8}
\end{equation}
Then define inner product as

\begin{equation}
\left\langle O_{1}\mid O_{2}\right\rangle \equiv\left\langle O_{1}O_{2}\right\rangle -\left\langle O_{1}\right\rangle \left\langle O_{2}\right\rangle =\left\langle \left(O_{1}-\overline{O_{1}}\right)\left(O_{2}-\overline{O_{2}}\right)\right\rangle ,\label{eq:F7}
\end{equation}
where $\overline{O}\equiv\left\langle O\right\rangle $.

Calculate $Q_{1}$ via $Q_{1}=\left\langle E_{C'}\right\rangle -\left\langle E_{D}\right\rangle +k_{B}T_{1}(\mathrm{ln}Z_{C}-\mathrm{ln}Z_{D})$
and $Q_{2}$ via $Q_{2}=\left\langle E_{B}\right\rangle -\left\langle E_{A'}\right\rangle +k_{B}T_{2}(\mathrm{ln}Z_{B}-\mathrm{ln}Z_{A})$
to $\mathcal{O}(\alpha^{2})$. We get

\begin{eqnarray}
Q_{1} & \thickapprox & \left[\frac{\beta_{2}}{\beta_{1}}\lambda_{B}\left\langle f\right\rangle _{B}-\lambda_{D}\left\langle f\right\rangle _{D}+\frac{1}{\beta_{1}}\left(\mathrm{ln}Z_{B}^{0}-\mathrm{ln}Z_{D}^{0}\right)\right]\nonumber \\
 &  & +\alpha\left[-\frac{\beta_{2}^{2}}{\beta_{1}}\lambda_{B}\left\langle f\mid g\right\rangle _{B}+\beta_{1}\lambda_{D}\left\langle f\mid g\right\rangle _{D}\right]\nonumber \\
 &  & +\alpha^{2}\left[-\beta_{2}\left\langle g\mid g\right\rangle _{B}+\frac{1}{2}\beta_{1}\left(\left\langle g\mid g\right\rangle _{B}+\left\langle g\mid g\right\rangle _{D}\right)\right]\nonumber \\
 &  & +\alpha^{2}\left[-\frac{\beta_{2}^{3}}{\beta_{1}}\lambda_{B}\left(\left\langle g\right\rangle _{B}\left\langle f\mid g\right\rangle _{B}-\frac{1}{2}\left\langle f\mid g^{2}\right\rangle _{B}\right)+\beta_{1}^{2}\lambda_{D}\left(\left\langle g\right\rangle _{D}\left\langle f\mid g\right\rangle _{D}-\frac{1}{2}\left\langle f\mid g^{2}\right\rangle _{D}\right)\right]\nonumber \\
 &  & +\alpha^{2}\left[-\left(\beta_{2}-\beta_{1}\right)\lambda_{C}^{1}\left\langle f\mid g\right\rangle _{B}+\frac{1}{2}\beta_{1}(\lambda_{C}^{1})^{2}\left\langle f\mid f\right\rangle _{B}\right],\label{eq:E27}
\end{eqnarray}

\begin{eqnarray}
Q_{2} & \thickapprox & \left[\lambda_{B}\left\langle f\right\rangle _{B}-\frac{\beta_{1}}{\beta_{2}}\lambda_{D}\left\langle f\right\rangle _{D}+\frac{1}{\beta_{2}}\left(\mathrm{ln}Z_{B}^{0}-\mathrm{ln}Z_{D}^{0}\right)\right]\nonumber \\
 &  & +\alpha\left[-\beta_{2}\lambda_{B}\left\langle f\mid g\right\rangle _{B}+\frac{\beta_{1}^{2}}{\beta_{2}}\lambda_{D}\left\langle f\mid g\right\rangle _{D}\right]\nonumber \\
 &  & +\alpha^{2}\left[-\frac{1}{2}\beta_{2}\left(\left\langle g\mid g\right\rangle _{B}+\left\langle g\mid g\right\rangle _{D}\right)+\beta_{1}\left\langle g\mid g\right\rangle _{D}\right]\nonumber \\
 &  & +\alpha^{2}\left[-\beta_{2}^{2}\lambda_{B}\left(\left\langle g\right\rangle _{B}\left\langle f\mid g\right\rangle _{B}-\frac{1}{2}\left\langle f\mid g^{2}\right\rangle _{B}\right)+\frac{\beta_{1}^{3}}{\beta_{2}}\lambda_{D}\left(\left\langle g\right\rangle _{D}\left\langle f\mid g\right\rangle _{D}-\frac{1}{2}\left\langle f\mid g^{2}\right\rangle _{D}\right)\right]\nonumber \\
 &  & +\alpha^{2}\left[-\left(\beta_{2}-\beta_{1}\right)\lambda_{A}^{1}\left\langle f\mid g\right\rangle _{D}-\frac{1}{2}\beta_{2}(\lambda_{A}^{1})^{2}\left\langle f\mid f\right\rangle _{D}\right].\label{eq:E28}
\end{eqnarray}
We can see that the variations of $\lambda_{C}$ and $\lambda_{A}$
don't alter $Q_{1}$ and $Q_{2}$ until $\mathcal{O}(\alpha^{2})$,
which implies that $\eta=1-T_{1}/T_{2}$ up to $\mathcal{O}(\alpha)$.
The feature will be discussed in the following.

\section{Analysis of efficiency and its optimization}

Now let's analyze $\eta$ order by order.

\subsection{Zeroth order}

By Eq.(\ref{eq:E27}) and Eq.(\ref{eq:E28}), the zeroth order term
of the efficiency is given by
\begin{equation}
\eta=1-\frac{T_{1}}{T_{2}},\label{eq:E29}
\end{equation}
as expected.

\subsection{First order}

Keep the efficiency to $\mathcal{O}(\alpha)$ and we have
\begin{equation}
\eta=1-\frac{T_{1}}{T_{2}}.\label{eq:E30}
\end{equation}
The first order term also gives Carnot efficiency. Why?

Consider that state $G$, which is almost at thermal equilibrium with
temperature $T$, undergoes a thermalization process by contacting
with a reservoir of temperature $T$, and then reaches state $H$.
We have $P_{n}(G)=P_{n}(H)+\delta P_{n},$ where $\delta P_{n}$ is
small. Calculate $\Delta S_{G\rightarrow H}^{total}$ to the second
order of $\delta P_{n}$ and we obtain
\begin{equation}
\Delta S_{G\rightarrow H}^{total}\thickapprox\frac{1}{2}k_{B}\sum_{n}\frac{\left(\delta P_{n}\right)^{2}}{P_{n}(H)}.\label{eq:F6}
\end{equation}

This is exactly what the second law of thermodynamics tells us. Entropy
has maximum at thermal equilibrium, which is state $H$ in our case,
and hence its first order expansion due to small departure vanishes.
Furthermore, for a spontaneous process, the total entropy cannot decrease,
and hence $\Delta S_{G\rightarrow H}^{total}$ is positive definite.

Due to the second law of thermodynamics, both $\Delta S_{C'\rightarrow C}^{total}$
and $\Delta S_{A'\rightarrow A}^{total}$ have vanishing first order
terms. Therefore, there is no efficiency correction up to $\mathcal{O}(\alpha)$
(by Eq.(\ref{eq:C8})).

\subsection{Second order}

Now let's turn to the second order term. Minimize $Q_{1}$ with $\lambda_{C}^{1}$
and maximize $Q_{2}$ with $\lambda_{A}^{1}$, and we get the maximum
of the efficiency
\begin{equation}
\eta=\left(1-\frac{T_{1}}{T_{2}}\right)+\alpha^{2}\left[-\frac{1}{2}\frac{\beta_{2}}{\beta_{1}}\left(\beta_{1}-\beta_{2}\right)^{2}\right]\frac{\left[\left\langle g\mid g\right\rangle -\frac{\left\langle f\mid g\right\rangle ^{2}}{\left\langle f\mid f\right\rangle }\right]_{B}+\left[\left\langle g\mid g\right\rangle -\frac{\left\langle f\mid g\right\rangle ^{2}}{\left\langle f\mid f\right\rangle }\right]_{D}}{\beta_{2}\lambda_{B}\left\langle f\right\rangle _{B}-\beta_{1}\lambda_{D}\left\langle f\right\rangle _{D}+\left(\mathrm{ln}Z_{B}^{0}-\mathrm{ln}Z_{D}^{0}\right)}.\label{eq:F21}
\end{equation}
This is the final result of the efficiency of the six-stroke QCE under
optimization with respect to $\lambda_{C}$ and $\lambda_{A}$.

If the physics makes sense, we should be able to prove mathematically
that $\eta$ is always not bigger than $1-T_{1}/T_{2}$. Here is the
proof. First, $\beta_{2}Q_{2}^{0}$ is positive and hence $\beta_{2}\lambda_{B}\left\langle f\right\rangle _{B}-\beta_{1}\lambda_{D}\left\langle f\right\rangle _{D}+\left(\mathrm{ln}Z_{B}^{0}-\mathrm{ln}Z_{D}^{0}\right)$
is positive. Second, by Cauchy-Schwarz inequality, $\left\langle g\mid g\right\rangle \left\langle f\mid f\right\rangle -\left\langle f\mid g\right\rangle ^{2}\geqslant0$.
Therefore, the efficiency is always not bigger than $1-T_{1}/T_{2}$.

In the following, we have four observations for the second order correction
of the efficiency. These tendencies can be seen in the numerical example
of \cite{xiao2015construction} (Figure 2 in \cite{xiao2015construction}).

\paragraph*{Observation 1:}

In our setting, $\lambda_{D}$ should be bigger than $\lambda_{C}^{0}$
in order to guarantee $Q_{1}>0$, which can be seen from Fig. 1. By
Eq.(\ref{eq:D15}), we have

\begin{equation}
\beta_{2}\lambda_{B}<\beta_{1}\lambda_{D}.\label{eq:G2}
\end{equation}
It is unphysical considering $\left(\lambda_{B},\lambda_{D}\right)$
with $\beta_{2}\lambda_{B}>\beta_{1}\lambda_{D}$.

\paragraph*{Observation 2:}

For $\beta_{2}\lambda_{B}\thickapprox\beta_{1}\lambda_{D}$, the isothermal
processes are short, and hence the nonequilibrium effect of the relaxation
processes becomes significant. For a fixed $\beta_{1}\lambda_{D}$,
the efficiency decreases sharply as $\beta_{2}\lambda_{B}$ approaching
its maximum ($\beta_{1}\lambda_{D}$); for a fixed $\beta_{2}\lambda_{B}$,
the efficiency decreases sharply as $\beta_{1}\lambda_{D}$ approaching
its minimum ($\beta_{2}\lambda_{B}$).

\paragraph*{Observation 3:}

Consider the limiting case of

\begin{equation}
\beta\lambda\gg1,\label{eq:G4}
\end{equation}
which corresponds to low temperature or high energy scale. Consider
only the contributions of the two lowest energy levels (two-level
system) as an approximation, and then we get

\begin{equation}
\left\langle g\mid g\right\rangle -\frac{\left\langle f\mid g\right\rangle ^{2}}{\left\langle f\mid f\right\rangle }\thickapprox0.\label{eq:G12}
\end{equation}
Take Eq.(\ref{eq:G12}) into Eq.(\ref{eq:F21}), and we can see that
efficiency correction goes to zero for $\beta\lambda\gg1$. This tendency
can be understood by the fact that a two-level system can always be
regarded as being at thermal equilibrium.

\paragraph*{Observation 4:}

Consider the limiting case of

\begin{equation}
\beta\lambda\ll1,\label{eq:G13}
\end{equation}
which corresponds to high temperature or low energy scale. In this
case, $P_{n}$ approaches a constant as $\beta\lambda$ goes to zero.
So efficiency correction also approaches a constant as $\beta\lambda$
goes to zero.

\section{Conclusions}

We analyze some properties of the six-stroke QCE. The features of
the six-stroke QCE come from the replacement of thermodynamic adiabatic
process with quantum mechanical adiabatic process. It is a non-trivial
step, and the meaning of such replacement is still an open question.
We hope that we can have conceptual understanding of such QCE and
quantum thermodynamics in the future.

\section{Acknowledgment}

This work is done under the advisement of Professor Yih-Yuh Chen,
my master program advisor. I gratefully thank him for insightful discussions.\medskip{}

\medskip{}
\medskip{}

\medskip{}
\medskip{}
\medskip{}

\medskip{}
\medskip{}

\medskip{}
\medskip{}

\bibliographystyle{plain}
\bibliography{thesis}

\end{document}